\documentclass[12pt,epsf,amstex]{article}
\usepackage [dvips]{graphicx}
\usepackage{amsmath}
\usepackage{amssymb}
\usepackage{epsfig}

\addtocounter{secnumdepth}{1}
\begin{document}
\newcommand{\br}{\bar{r}}
\newcommand{\bbeta}{\bar{\beta}}
\newcommand{\bgamma}{\bar{\gamma}}
\newcommand{\tbeta}{\tilde{\beta}}
\newcommand{\tgamma}{\tilde{\gamma}}
\newcommand{\bE}{{\bf{E}}}
\newcommand{\bO}{{\bf{O}}}
\newcommand{\bR}{{\bf{R}}}
\newcommand{\bS}{{\bf{S}}}
\newcommand{\bT}{\mbox{\bf T}}
\newcommand{\bt}{\mbox{\bf t}}
\newcommand{\half}{\frac{1}{2}}
\newcommand{\summ}{\sum_{m=1}^n}
\newcommand{\sumq}{\sum_{q=1}^\infty}
\newcommand{\sumqno}{\sum_{q\neq 0}}
\newcommand{\prodm}{\prod_{m=1}^n}
\newcommand{\prodq}{\prod_{q=1}^\infty}
\newcommand{\maxm}{\max_{1\leq m\leq n}}
\newcommand{\minm}{\min_{1\leq m\leq n}}
\newcommand{\maxphi}{\max_{0\leq\phi\leq 2\pi}}
\newcommand{\tsum}{\Sigma}
\newcommand{\bsA}{\mathbf{A}}
\newcommand{\bsV}{\mathbf{V}}
\newcommand{\bsE}{\mathbf{E}}
\newcommand{\bsT}{\mathbf{T}}
\newcommand{\bsZ}{\hat{\mathbf{Z}}}
\newcommand{\bse}{\mbox{\bf{1}}}
\newcommand{\bspsi}{\hat{\boldsymbol{\psi}}}
\newcommand{\cdottt}{\!\cdot\!}
\newcommand{\deltaR}{\delta\mspace{-1.5mu}R}
\newcommand{\invup}{\rule{0ex}{2ex}}

\newcommand{\bGamma}{\boldmath$\Gamma$\unboldmath}
\newcommand{\dd}{\mbox{d}}
\newcommand{\ee}{\mbox{e}}
\newcommand{\p}{\partial}
\newcommand{\expmVo}{\langle\ee^{-{\mathbb V}}\rangle_0}

\newcommand{\Rav}{R_{\rm av}}
\newcommand{\dav}{d_{\rm av}}
\newcommand{\Rc}{R_{\rm c}}
\newcommand{\rmax}{r_{\rm max}}

\newcommand{\la}{\langle}
\newcommand{\ra}{\rangle}
\newcommand{\rao}{\rangle\raisebox{-.5ex}{$\!{}_0$}}  
\newcommand{\rae}{\rangle\raisebox{-.5ex}{$\!{}_1$}}
\newcommand{\raG}{\rangle_{_{\!G}}}
\newcommand{\rainr}{\rangle_r^{\rm in}}
\newcommand{\beq}{\begin{equation}}
\newcommand{\eeq}{\end{equation}}
\newcommand{\bea}{\begin{eqnarray}}
\newcommand{\eea}{\end{eqnarray}}
\def\lsim{\:\raisebox{-0.5ex}{$\stackrel{\textstyle<}{\sim}$}\:}
\def\gsim{\:\raisebox{-0.5ex}{$\stackrel{\textstyle>}{\sim}$}\:}

\numberwithin{equation}{section}

\thispagestyle{empty}
\title{\Large {\bf 
Sylvester's question\\[3mm]
and the Random Acceleration Process
\phantom{xxx} }}
 
\author{{H.\,J. Hilhorst$^1$, P. Calka$^2$, and G. Schehr$^1$}\\[5mm]
{\small $^1$Laboratoire de Physique Th\'eorique, B\^atiment 210}\\[-1mm] 
{\small Univ Paris-Sud 11 and CNRS,
91405 Orsay Cedex, France}\\
{\small $^2$Laboratoire MAP5, Universit\'e Paris Descartes, 45, rue des
  Saints-P\`eres}\\[-1mm] 
{\small 75270 Paris Cedex 06}\\}

\maketitle
\begin{small}
\begin{abstract}
\noindent 
Let $n$ points be chosen randomly and independently in the unit disk.
``Sylvester's question'' concerns 
the probability $p_n$ that they are the vertices of a
convex $n$-sided polygon. 
Here we establish the link with another problem.
We show that for large $n$ this polygon, when suitably parametrized by 
a function ${r}(\phi)$ of the polar angle $\phi$, 
satisfies the equation of the random acceleration process (RAP),
$\dd^2{r}/\dd\phi^2 = f(\phi)$, 
where $f$ is Gaussian noise. 
On the basis of this relation we derive the asymptotic expansion
$\log p_n = -2n\log n + n\log(2\pi^2\ee^2) - c_0 n^{1/5} + \ldots$, of which
the first two terms agree with a rigorous 
result due to B\'ar\'any. The nonanalyticity in $n$ of the third term 
is a new result. 
The value $\frac{1}{5}$ of the exponent follows from
recent work on the RAP due to Gy\"orgyi 
{\it et al.}~[{\it Phys.~Rev.~E\,} {\bf 75}, 021123 (2007)].
We show that
the $n$-sided polygon is effectively contained in an annulus of width 
$\,\sim n^{-4/5}$ along the edge of the disk. 
The distance $\delta_n$ of closest approach to the edge 
is exponentially distributed with average $(2n)^{-1}$.
\\

\noindent
{{\bf Keywords:} random convex polygons, random points in convex position,
random acceleration process, integrated Brownian motion}
\end{abstract}
\end{small}
\vspace{12mm}

\noindent LPT Orsay 08/61
\newpage


\section{Introduction} 
\label{secintroduction}

\begin{figure}
\begin{center}
\scalebox{.55}
{\includegraphics{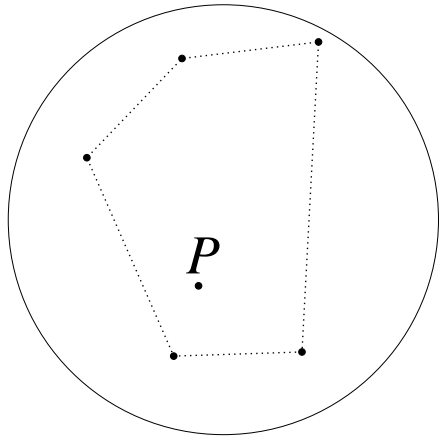}}
\end{center}
\caption{\small An instance of a set of six 
points chosen randomly in a disk $D$.
The convex envelope (dotted) of this set 
does not pass through
the point $P$ and hence is a five-sided polygon.
This set of points therefore does not contribute to $p_6(D)$.}
\label{figdisk}
\end{figure}

Let $n$ points be chosen independently according to a uniform distribution 
on a disk $D \subset {\mathbb R}^2$.
We consider in this work the probability $p_n(D)$
that these points are the vertices of a convex $n$-sided polygon; 
an example is shown in figure \ref{figdisk}.
This question,
with the disk $D$ replaced by an arbitrary convex
domain $K \subset {\mathbb R}^2$, has a long history in mathematics.
In 1864 Sylvester \cite{Sylvester1864} asked about the value of $p_4(K)$.

For a parallelogram $S$ and a triangle $T$ the exact results
\beq
p_n(S) = \Bigg[ \frac{1}{n!}\binom{2n-2}{n-1} \Bigg]^2, \qquad
p_n(T) = \frac{2^n(3n-3)!}{(n-1)!^{\,3}\,(2n)!} 
\label{exfinres}
\eeq
were shown by Valtr \cite{Valtr95,Valtr96}.
Since affine transformations may transform any parallelogram into a square and
any triangle into an equilateral triangle
while leaving the distribution uniform, 
expressions (\ref{exfinres})
do not depend on the particular choice of triangle or parallelogram.
In the limit of large $n$, the probabilities $p_n(K)$ become very
small and we will be interested in their asymptotic large-$n$ expansions.
Those of $p_n(S)$ and $p_n(T)$ above are given by
\bea
\log p_n(S) &=& -2n\log n + n\log(16\ee^2)  + {\cal O}(\log n), 
\nonumber \\[2mm]
\log p_n(T) &=& -2n\log n + n\log(\tfrac{27}{2}\ee^2) + {\cal O}(\log n), 
\label{asptST}
\eea
in which only the coefficients of the terms linear in $n$ are different.
These coefficients were subsequently explained by B\'ar\'any \cite{Barany99},
who derived a general result 
that we will paraphrase as follows.
B\'ar\'any showed that for an arbitrary convex domain $K$ of area $A_K$
\beq
\log p_n(K) = -2n\log n + 
n\log\Big( \tfrac{1}{4}\ee^2 \frac{{\cal P}_{K}^{\,3}}{A_K} \Big)
+ o(n),
\label{asptK}
\eeq
in which the meaning of ${\cal P}_K$ still has to be given. 
To do so we need the concept of the {\it affine length\,} 
${\cal L}_C$ of a convex curve $C$, defined (see {\it e.g.} \cite{Barany99}) as
\beq
{\cal L}_C = \int_C \kappa^{1/3}\, \dd s,
\label{defaffper}
\eeq
with $\,\dd s\,$ a line element along the curve and $\kappa$ the
local inverse radius of curvature.
We note that the affine length has the physical dimensionality
of a (length)${}^{{2}/{3}}$.
Let now a convex subdomain $K'$ of $K$ have a border $\partial K'$ 
of affine length ${\cal L}_{\partial K'}$
and let furthermore ${\cal L}_{\partial K'}$ attain its
maximum for $K'=K'_{\rm max}$. Then
\beq
{\cal P}_{K} \equiv \max_{K'\subseteq K} {\cal L}_{\partial K'}
= {\cal L}_{\partial K'_{\rm max}}
\label{defcalP}
\eeq
is called the {\it affine perimeter\,} of $K$.
This completes the definition of (\ref{asptK}).

There is no simple way to calculate ${\cal P}_{K}$ for general $K$.
When $K$ is a disk $D$ of radius $R_D$, symmetry dictates that 
$K'_{\rm max}=K=D$. Then (\ref{defcalP}) and 
(\ref{defaffper}) with $\kappa=R_D^{-1}$ yield 
${\cal P}_K = {\cal L}_{\partial D} = 2\pi R_D^{2/3}$, 
so that (\ref{asptK}) becomes
\beq
\log p_n(D) = -2n\log n + n\log(2\pi^2\ee^2) + {\cal R}_n\,,
\label{asptD}
\eeq
in which the remainder ${\cal R}_n$ is $o(n)$. This statement 
about the remainder is weaker than
that in equation (\ref{asptST}), 
where the rest term is known to be ${\cal O}(\log n)$.
It will appear that there is a good reason for this difference. 
\vspace{3mm}

In this work we represent the $n$ random points by polar
coordinates $(R_m,\Phi_m)$ for $m=1,\ldots,n$ in such a way that
$0<\Phi_1<\Phi_2<\ldots<\Phi_n \leq 2\pi$.
We define the average radius, $\Rav$, and the scaled 
deviation from average, $r_m$, by 
\beq
\Rav \equiv n^{-1}\summ R_m\,, \qquad R_m=\Rav(1+n^{-1/2}r_m).
\label{defRavrm}
\eeq
For $n\to\infty$ with $\phi\equiv 2\pi mn^{-1}$ fixed,
$r_m$ becomes a random function $r(\phi)$ on $[0,2\pi]$.
Our principal result
is that in this limit the remainder ${\cal R}_n$ in (\ref{asptD})
can be cast in the form 
\beq
{\cal R}_n = \log \big\la \exp\big[ -2n^{{1}/{2}}\maxphi r(\phi) 
             \big]\big\ra_0\,,
\label{exprcalRn}
\eeq
in which $\la\ldots\ra_0$ denotes the average with respect to all
$2\pi$-periodic zero-integral 
solutions $r(\phi)$ of 
\beq
\frac{\dd^2 r(\phi)}{\dd\phi^2} = f(\phi),
\label{introracp}
\eeq
where $f(\phi)$ is Gaussian noise of autocorrelation
\beq
\la f(\phi)f(\phi') \ra_0 = 
\tfrac{3}{2}\big[ 2\pi\delta(\phi-\phi')\,-\, 1 \big].
\label{exprautocorrintro}
\eeq
We arrive at these results
by extending an analytic method that was developed originally
in the context of planar Voronoi tessellations
\cite{Hilhorst05a,Hilhorst05b,Hilhorst08a}
and employed also to study 
Poisson line tessellations in general and the Crofton cell
problem in particular \cite{HilhorstCalka08}.
The idea of applying this method here arose from the observation
that the first two terms in expansion (\ref{asptD}) are 
identical to those of the expansion of $\log p_n^{\rm Vor}$,
where $p_n^{\rm Vor}$ is the probability for the typical Poisson-Voronoi cell
to be $n$-sided. The work of reference \cite{CalkaSchreiber05} suggests that
the two problems are related by an inversion of the radial coordinates with
respect to the unit circle; this is indeed borne out by our analysis.
Our method consists of a rather intricate but fully exact coordinate
transformation, followed by an asymptotic expansion which, 
although nonrigorous, 
is of the kind routinely used in physics to obtain exact results.

Equation (\ref{introracp}) is known as the {\it random acceleration process;} 
the function $r(\phi)$ is also referred to as {\it Kolmogorov diffusion} or
{\it integrated Brownian motion}.
We will briefly review some of the literature on this stochastic process
in section \ref{secproperties}. 
Of particular relevance to us is recent work by Gy\"orgyi {\it et al.} 
(GMOR) \cite{GMOR07}.  
Upon using their result for the right hand side of (\ref{exprcalRn}) 
we find that (\ref{asptD}) becomes
\beq
\log p_n(D) = -2n\log n + n\log(2\pi^2\ee^2) 
- 2\epsilon_0 (3\pi^4 n)^{1/5} + \ldots,
\label{mainresult}
\eeq
where $\epsilon_0>0$ is
the smallest eigenvalue of a linear eigenvalue problem,
the only hypothesis being the existence of its solution.
An immediate corollary is that in the large-$n$ limit
the average (\ref{exprcalRn}) draws its main contribution from polygons
that stay within a distance
of order $\sim n^{-4/5}R_D$ from the edge of the disk $D$.
For $n\to\infty$ 
the distance of closest approach to the disk is shown to be exponentially
distributed with average $1/(2n)$.

The term proportional to $n^{1/5}$ in (\ref{mainresult})
is a new contribution to the answer to Sylvester's question. 
Since such a nonanalytic term is absent from the expansions (\ref{asptST}) 
for the triangle and the square,
we are led to ask under which conditions
such an $n^{1/5}$ term appears. B\'ar\'any \cite{Barany99}
showed, essentially, that in the large-$n$ limit the $n$ points in convex
position lie on the curve $\partial K'_{\rm max}$,
and that this curve is composed of 

${}$\phantom{i}(i) 
arcs or isolated points that coincide with the 
domain boundary $\partial K$;

(ii) arcs of parabolas in the interior of $K$.\\
The present study provides strong indication
that an $n^{1/5}$ term occurs whenever $\partial K'_{\rm max}$ 
contains at least one arc coinciding with $\partial K$,
that is, when $\partial K'_{\rm max}$ sticks to the
domain boundary over some nonzero angular interval.
This is obviously the case for the circle, where
$\partial K'_{\rm max} = \partial K$, but not for the square and the triangle,
where the limit curve $\partial K'_{\rm max}$ touches the domain boundary 
$\partial K$ only in isolated points \cite{Valtr95,Valtr96}. 

In section \ref{secdisk} we carry out the exact coordinate transformation.
In section \ref{secracp} we perform the large-$n$ expansion and establish 
relation (\ref{exprcalRn}).
In section \ref{secGMOR} we use GMOR's results to obtain
(\ref{mainresult}).
In section \ref{secheuristic}
we show how to obtain the power $\frac{1}{5}$ in (\ref{mainresult}) 
by heuristic arguments.
In section \ref{secweaker} we return to the hypothesis of the existence of the
eigenvalue $\epsilon_0$; we derive exact bounds for the right hand side of 
(\ref{exprcalRn}) and show that even without
the existence hypothesis the conclusion remains valid that ${\cal R}_n$ 
is nonanalytic in $n$.
Section \ref{secapproach} contains the 
study of the distribution of the distance of closest approach of the edge.
Section \ref{secconclusion} is our conclusion.


\section{Random convex polygon in a disk}
\label{secdisk}

We consider $n$ points $\bR_1,\ldots,\bR_n$
drawn randomly and independently from a uniform
distribution on the disk of radius $R_D$, centered in the origin.
We ask for the probability $p_n(D)$ that 
these $n$ points are the vertices of a convex polygon.
A slightly different probability $p_n^*$
is defined the same way but with the additional condition 
that the polygon enclose the origin. 
When $n$ gets large, the ratio between $p_n(D)$ and $p^*_n$ will 
tend to unity exponentially rapidly with $n$ \cite{footnote0}, 
but $p_n^*$ will be easier to study.
We begin by writing its definition.
In terms of the polar coordinate representation $\bR_m=(R_m,\Phi_m)$ we have 
\beq
p^*_n = \frac{1}{(\pi R_D^2)^{n}_{\invup}} 
\int_0^{R_D} \!R_1\dd R_1\ldots R_n\dd R_n
\int_0^{2\pi} \!\dd\Phi_1\ldots \dd\Phi_n \,\chi(\bR_1,\ldots,\bR_n),
\label{expn}
\eeq 
in which $\chi$ is the indicator of the subdomain of phase
space where the points form an origin-enclosing convex $n$-sided polygon;
the explicit expression of $\chi$ will be discussed in section \ref{secchi}.
Everywhere below we will scale the `radii' $R_m$ such that $R_D=1$.

We will now subject expression (\ref{expn}) to a series of coordinate
transformations. 
The final result of these will be
equation (\ref{expnV}) together with (\ref{defmathbbV}) and (\ref{defP0}).


\subsection{Transformation of variables. I}
\label{sectransfI}

In (\ref{expn}) we may set one of the angles, say $\Phi_n$, equal to $2\pi$
if we compensate by an extra factor $2\pi$;
and the remaining $n-1$ angles may be ordered such that
\beq
0<\Phi_1<\ldots<\Phi_{n-1}<\Phi_n = 2\pi
\label{orderPhim}
\eeq
if a compensating factor $(n-1)!$ is introduced.
Referring now to figure \ref{figconvex} we define the 
angle differences $\xi_m$ between two consecutive vertex 
vectors $\bR_{m-1}$ and $\bR_m$ by
\beq
\xi_m = \Phi_m-\Phi_{m-1}, \qquad m=1,\ldots,n,
\label{defxim}
\eeq
with the convention $\Phi_0=0$. Then the $\xi_m$ are positive 
and satisfy the sum rule $\summ\xi_m=2\pi$.
In terms of these variables we can write (\ref{expn}) as
\beq
p^*_n=\frac{2\pi(n-1)!}{\pi^{n}}
\int_0^{1} \! R_1\dd R_1\ldots R_n\dd R_n
\int_0^{2\pi}\! \dd\xi_1\ldots\dd\xi_n\, \delta(\xi_1+\ldots+\xi_n-2\pi)\,\chi.
\label{expnRxi}
\eeq
Next we define the average radius $\Rav$ and
the reduced radii $\rho_m$ by 
\beq
\Rav=\frac{1}{n}\sum_{m=1}^n R_m, \qquad R_m=\rho_m\Rav.
\label{defRavrhom}
\eeq
The $\rho_m$ therefore satisfy the sum rule
\beq
\frac{1}{n}\summ\rho_m = 1.
\label{sumrho}
\eeq
We now rewrite the integrals on the $R_m$ in (\ref{expnRxi}) as
\bea
\int_0^{1} \!\dd R_1\ldots\dd R_n &=&
\int_0^{1} \!\dd\Rav\,\Rav^{2n-1}\int_0^{\Rav^{-1}} 
\!\dd\rho_1\ldots\dd\rho_n\,
\delta\Big( 1- n^{-1}\summ\rho_m \Big)
\nonumber\\[2mm]
&=&
\int_0^{\infty} \!\dd\rho_1\ldots\dd\rho_n\,
\delta\Big( 1- n^{-1}\summ\rho_m \Big)\, \nonumber\\[2mm]
&& \times
\int_0^{1} \!\dd\Rav\,\Rav^{2n-1} \prodm \theta\big( \Rav^{-1}-\rho_m \big)
\label{transfRint}
\eea
where  $\theta\big( \Rav^{-1}-\rho_m \big)$ is the Heaviside step function, equal to $1$ if $\Rav^{-1}-\rho_m>0$ and to $0$ if  $\Rav^{-1}-\rho_m<0$. If (\ref{transfRint}) is applied to an integrand without $\Rav$ dependence,
then the integration on $\Rav$ may be carried out according to
\bea
\int_0^1\!\dd\Rav\, \Rav^{2n-1} \prodm\theta(\Rav^{-1}-\rho_m)
&=& \!\int_0^{\min_{1\leq m\leq n} \rho_m^{-1}} \dd\Rav \Rav^{2n-1}
\nonumber\\[2mm]
&=& (2n)^{-1}\Big[ \maxm \rho_m \Big]^{-2n}.
\label{intonRav}
\eea
Hence, after we substitute (\ref{intonRav}) in (\ref{transfRint}) and
(\ref{transfRint}) in (\ref{expnRxi}) we get
\bea
p^*_n &=& \frac{(n-1)!}{n\pi^{n-1}} 
\int_0^{2\pi}\!\dd\xi_1\ldots\dd\xi_n\,\delta\Big( \summ\xi_m-2\pi \Big)
\nonumber\\[2mm]
&& \times
\int_0^\infty\dd\rho_1\ldots\dd\rho_n\,
\delta \Big( 1-n^{-1}\summ\rho_m \Big)\,
\chi\, \Big[ \maxm \rho_m \Big]^{-2n}.
\label{intRrho}
\eea
This completes the conversion of the variables of integration from the $R_m$
and $\Phi_m$ to the $\rho_m$ and $\xi_m$.


\subsection{The convexity condition}
\label{secchi}

\begin{figure}
\begin{center}
\scalebox{.55}
{\includegraphics{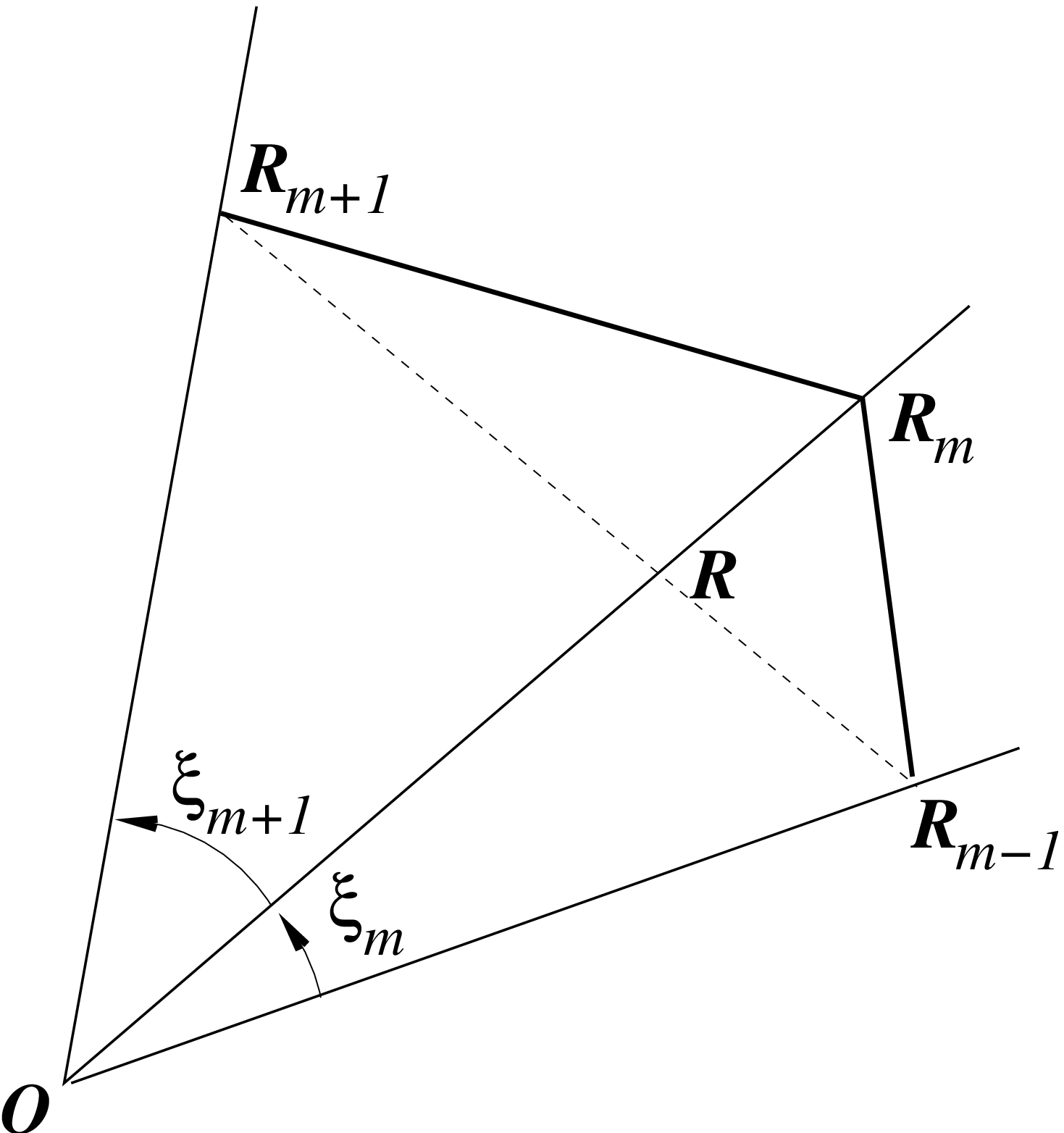}}
\end{center}
\caption{\small
Figure illustrating the derivation of the convexity condition 
(\ref{xirhocond}).}
\label{figconvex}
\end{figure}

We will now find an explicit expression for the convexity condition,
enforced by the indicator $\chi$,
in terms of the variables of integration $\rho_m$ and $\xi_m$.
This is most easily done by means of the following
geometric argument. In figure \ref{figconvex}, 
the point $\bR$ is the intersection of the line segments
$\bO\bR_m$ and $\bR_{m-1}\bR_{m+1}$.
The point $\bR_m$ is in convex position with respect to $\bR_{m-1}$ and
$\bR_{m+1}$ if $R<R_m$,
or equivalently, if
\beq
\mbox{area}(\bO\bR_{m-1}\bR_m) + \mbox{area}(\bO\bR_{m+1}\bR_m) > 
\mbox{area}(\bO\bR_{m-1}\bR_{m+1}).
\label{compareas}
\eeq
The areas are readily expressed in terms of the two angles $\xi_m$ and
$\xi_{m+1}$ and the three radii $R_{m-1}$, $R_m$, and $R_{m+1}$.
After division by $R_{m-1}R_mR_{m+1}\Rav^{-1}$ 
the length scale drops out and (\ref{compareas}) becomes
\beq
\frac{\sin\xi_m}{\raisebox{0.6ex}{$\rho_{m+1}$}} + 
\frac{\sin\xi_{m+1}}{\raisebox{0.6ex}{$\rho_{m-1}$}} >
\frac{\sin(\xi_m+\xi_{m+1})}{\raisebox{0.6ex}{$\rho_m$}}\,, 
\qquad m=1,\ldots,n,
\label{xirhocond}
\eeq
where $\xi_m$ and $\rho_m$ are understood to be $n$-periodic in their index. 
A similar condition, differing only in that
all radii are replaced by their inverses,
was encountered in the case of the
Voronoi cells \cite{Calka03b,Hilhorst05b}; it there expresses the condition
that the $m$\,th point (for $m=1,\ldots,n$) contribute a nonzero segment to the
perimeter of the Voronoi cell. This property is due to a duality argument:
when $C$ is a convex set containing the origin in its interior, its dual (or
polar body) $C^*$
is the convex set $\{x\in{\mathbb R}^2;\la x,y\ra \le 1 \;\;
\forall\;y\in C\}$ where
$\la\cdot,\cdot\ra$ is the usual scalar product (see
\cite{Schneider93}). Formally, when applied to a convex $n$-sided
polygon, the transformation provides a new convex $n$-sided polygon such that 
the projections of the origin onto its edges are given by the polar
coordinates $(R_m^{-1},\Phi_m)$, $1\le m\le n$. 

Equations (\ref{intRrho}) and (\ref{xirhocond}) represent an intermediate
stage in the transformation of variables.
Equation (\ref{xirhocond}) couples the variables of integration nontrivially
and makes the problem of evaluating (\ref{intRrho}) a hard one.


\subsection{New angular variables}
\label{secangular}

\begin{figure}
\begin{center}
\scalebox{.55}
{\includegraphics{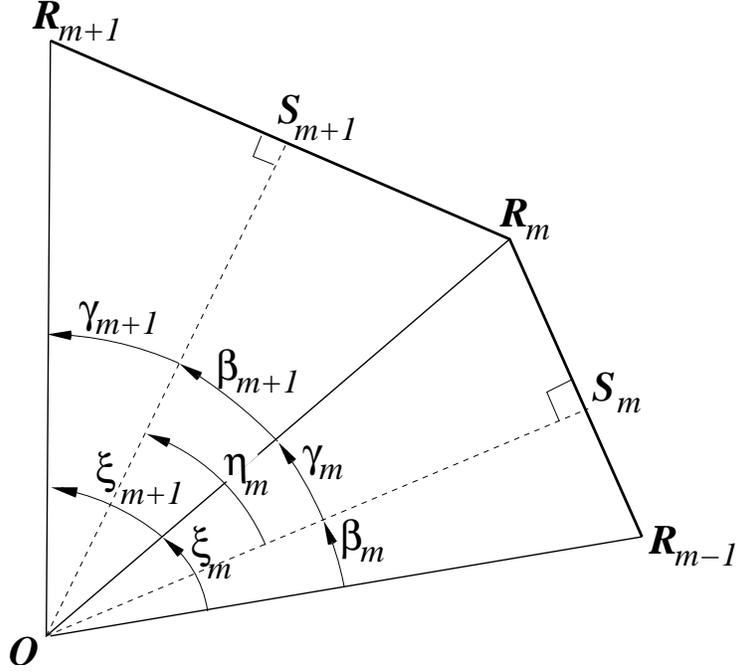}}
\end{center}
\caption{\small
Figure illustrating the definition of the angles 
$\xi_m,\, \eta_m,\, \beta_m$, and $\gamma_m$.}
\label{figangles}
\end{figure}

In this subsection we introduce the angular variables necessary for the
remainder of our analysis. We refer to figure \ref{figangles}.

Let $\bS_m=(S_m,\Psi_m)$ be the projection of the origin onto the 
line passing through $\bR_{m-1}$ and $\bR_m$; the $S_m$ and $\Psi_m$ may be
expressed in terms of the $R_m$ and $\Phi_m$.
The remaining angles needed in the discussion  
may then be defined in terms of the $\Phi_m$ and $\Psi_m$. 
The angle difference $\eta_m$
between the projection vectors $\bS_{m}$ and $\bS_{m+1}$ is
\beq
\eta_m = \Psi_{m+1}-\Psi_m, \qquad m=1,2,\ldots,n,
\label{defeta}
\eeq
with the convention $\Psi_{n+1}=\Psi_1+2\pi$.
The $\eta_m$ satisfy the sum rule $\summ\eta_m=2\pi$. 
Next we define $2n$ angles between projection and vertex vectors,
\beq
\beta_m=\Psi_m-\Phi_{m-1}, \qquad \gamma_m=\Phi_m-\Psi_m, \qquad
m=1,\ldots,n.
\label{defbg}
\eeq
We note that 
the $\beta_m$ and $\gamma_m$ may be negative,
as happens when the projection $\bS_m$
falls outside the line segment connecting $\bR_{m-1}$ and $\bR_m$.
In any case the geometry shows that 
we must have 
\beq
-\frac{\pi}{2}<\beta_m,\gamma_m<\frac{\pi}{2}.
\label{ineqbg}
\eeq
For fixed sets of angles
$\xi=\{\xi_m\}$ and $\eta=\{\eta_m\}$ one may still jointly rotate
the vertex vectors $\bR_m$ with respect to the projection vectors
$\bS_m$,
as this modifies only the relative angles $\beta_m$ and $\gamma_m$ 
(see figure \ref{figangles}) between the two sets. 
We may select any one of these relative angles and call it `the' angle of
rotation, since it will determine all others. We will select $\beta_1$
as this special degree of freedom and express
the remaining $\beta_m$ and $\gamma_m$ as
\bea
\beta_m &=& \phantom{-}\beta_1 - 
\sum_{\ell=1}^{m-1}(\xi_{\ell}-\eta_\ell),
\qquad m=2,\ldots,n,
\nonumber\\
\gamma_m &=& -\beta_1 +
\sum_{\ell=1}^{m-1}(\xi_{\ell}-\eta_{\ell}) + \xi_m\,, 
\qquad m=1,\ldots,n,
\label{inversebgxy}
\eea
Now observe that the reduced radii $\rho_m$ satisfy the relations
\beq
\frac{\rho_{m-1}}{\rho_{m}} = \frac{\cos\gamma_m}{\cos\beta_m}\,, \quad
m=1,\ldots,n,
\label{ratiorho}
\eeq
where $\rho_0 \equiv \rho_n$.
This ratio (\ref{ratiorho}) is the same as the one encountered 
in the Voronoi and Crofton 
cell problems \cite{Hilhorst05b,HilhorstCalka08}, 
except for an 
{\it interchange of $\beta_m$ and $\gamma_m$}. 
It follows that one may express the 
$\rho_m$ exclusively in terms of the angles by solving them from  
(\ref{ratiorho}) together with (\ref{sumrho}).

We return now to the convexity condition (\ref{xirhocond}).
We may use (\ref{ratiorho}) to eliminate
$\rho_{m\pm 1}$ from (\ref{xirhocond}) in favor of $\rho_m$, 
which subsequently divides out.
After some trigonometry (\ref{xirhocond}) appears to reduce to
the condition\, $\tan\gamma_{m}+\tan\beta_{m+1}>0$.
Because of (\ref{ineqbg})
this is equivalent to $\gamma_m+\beta_{m+1} > 0$. Still using 
(\ref{inversebgxy}) we see that
\beq
\eta_m>0, \qquad m=1,2,\ldots,n,  
\label{condeta}
\eeq
is, finally, the condition for the $n$ points to be in convex position.


\subsection{Periodicity in the polar angle}
\label{secG}

Let us next define the function $G$ by
\beq
\ee^{2\pi G(\xi,\eta;\beta_1)} = 
\prod_{m=1}^n \frac{\cos\gamma_m}{\cos\beta_m}\,,
\label{defG}
\eeq
with the $\beta_m$ and $\gamma_m$ given by (\ref{inversebgxy}).
This function appears when we relate $\rho_n$ to $\rho_0$ by taking the
product of (\ref{ratiorho}) on $m=1,\ldots,n$.
Because of the periodicity condition $\rho_0\equiv\rho_n$ 
the angle of rotation $\beta_1$ cannot be arbitrary:
it must have the special value $\beta_1=\beta_*(\xi,\eta)$ 
that is the solution of
\beq
G(\xi,\eta;\beta_*)=0.
\label{nospiralcond}
\eeq
This equation is {\it invariant under the interchange of $\beta_m$ and
  $\gamma_m$} and is therefore {\it identical\,} 
to the one that appeared in the Voronoi and Crofton cell problems.
Indeed, as has already been said, the
dual body of a convex $n$-sided polygon is a $n$-sided polygon such that the
sets of angles $\{\xi_m\,;\,1\le m\le n\}$ and $\{\eta_m\,;\,1\le m\le n\}$ 
(as well as 
$\{\beta_m\,;\,1\le m\le n\}$ and $\{\gamma_m\,;\,1\le m\le n\}$) have been
exchanged. 
It was shown in reference \cite{Hilhorst07} that the solution 
$\beta_*(\xi,\eta)$ of (\ref{nospiralcond}), subject to (\ref{ineqbg}), 
exists and is unique
if and only if the $\xi_m$ and $\eta_m$ satisfy the condition
\beq
\max_{1\leq m\leq n} \Big[ \sum_{\ell=1}^{m-1}(\xi_\ell-\eta_\ell)\,+\,
\xi_m \Big]\, -\,
\min_{1\leq m\leq n} \Big[ \sum_{\ell=1}^{m-1}(\xi_\ell-\eta_\ell) \Big]\, 
<\, \pi.
\label{condxy}
\eeq
We will let $\Theta(\xi,\eta)$ denote the indicator function, equal to zero or
to unity, of the domain in phase space where (\ref{condxy}) is 
fulfilled.


\subsection{Transformation of variables. II}
\label{sectransfII}

Finally we carry out the transformation that eliminates
the $\rho_m$ occurring in (\ref{intRrho}) in favor of the $\eta_m$
defined in  (\ref{defeta}).
The details of this transformation step have been described in
\cite{Hilhorst05b} and the appendix of \cite{Hilhorst07}, and we will not
reproduce them here.
The result is an integral on the two sets of variables $\xi$ and $\eta$.
It is described conveniently with the aid
of a `zeroth order' probability distribution 
${\cal P}^{(0)}(\xi,\eta)$ defined as
\beq
{\cal P}^{(0)}(\xi,\eta) = \frac{1}{{\cal N}}
\Bigg[ \prodm\xi_m \Bigg]\,
\delta\Big(\summ\xi_m-2\pi\Big)\delta\Big(\summ\eta_m-2\pi\Big),
\label{defP0}
\eeq
where ${\cal N}=(2\pi)^{3n-2}/[(2n-1)!(n-1)!]$ 
is the appropriate normalization constant.
The average of any quantity $X$ with respect to 
${\cal P}^{(0)}$ will be denoted as $\la X\ra_0$. 
The transformation then leads to
\beq
p^*_n = p_n^{(0)}\la \Theta\ee^{-{\mathbb V}} \ra_0\,, \qquad
p_n^{(0)} = \frac{1}{4\pi^2}\,\frac{(8\pi^2)^n}{(2n)!}\,,
\label{expnV}
\eeq
in which $\Theta$ is defined following (\ref{condxy}) 
and ${\mathbb V}$ is given by 
\beq
\ee^{-{\mathbb V}} = \frac{1}{G'(\xi,\eta;\beta_*)}
\Bigg[ \prodm \frac{ \rho_m^2  \sin\xi_m }{ \xi_m\cos\beta_m\cos\gamma_m } 
\Bigg] 
\Big[ \max_{1\leq m\leq n}\rho_m \Big]^{-2n},
\label{defmathbbV}
\eeq
where the prime on $G$ denotes differentiation with respect to its last
argument. 
The factor $1/\xi_m$ in the product in (\ref{defmathbbV})
is compensated by the $\xi_m$ in the product in (\ref{defP0}).
This has been so arranged in order that $\exp(-{\mathbb V})$
do not diverge for any of the $\xi_m$ tending to zero.
We note in passing that $p_n^{(0)}$ is the same zeroth order
probability that occurs in the Voronoi problem \cite{Hilhorst05b,Hilhorst07}. 
Equations (\ref{defP0})-(\ref{defmathbbV}) are fully exact
and are at the basis of all further developments.


\section{Large-$n$ limit and Random Acceleration 
Process}
\label{secracp}

The initial problem (\ref{expn}) has been transformed into
the evaluation of the average $\la\Theta\ee^{-{\mathbb V}}\ra_0$ 
in (\ref{expnV}).
This, however, is still a formidable problem.
>From here on we will be able to proceed only by making a large-$n$ expansion.
This is the subject of this section.

The average involves only angles and 
we make the hypothesis, to be confirmed by the calculation, 
that for large $n$ they all scale with negative powers of $n$.
To begin with, 
we study the scaling that follows from ${\cal P}^{(0)}(\xi,\eta)$.
At a later stage we will then discuss how this scaling is 
modified by the presence of the integrand $\Theta\ee^{-{\mathbb V}}$
in (\ref{expnV}).


\subsection{Scaling in the large-$n$ limit. I}
\label{secscalingI}

We start with a preliminary.
We note that according to ${\cal P}^{(0)}$
the $\xi_m$ are independent apart from the sum rule constraint
represented by the delta function; and a similar statement
holds for the $\eta_m$.
Furthermore, the $\xi_m$ and $\eta_\ell$ are mutually fully independent.
More precisely, let $\{X_m;m\ge 1\}$  and $\{Y_l;l\ge 1\}$ be two independent
sequences of i.i.d.~random variables such that $Y_l$ is exponentially
distributed with mean 1 and $X_m$ is distributed with law
$u(x)=4x\ee^{-2x}$, $x>0$. For a fixed $n$, the variable $\xi_m$
(resp. $\eta_l$), $1\le m,l\le 
n$, is equal  in law to $(2\pi X_m)/(\sum_{k=1}^n X_k)$ (resp. to
$(2\pi Y_l)/(\sum_{k=1}^nY_k)$). Indeed, let us average
any positive bounded test function
$h(\eta_1,\ldots,\eta_n)$ on $[0,2\pi]^n$
with respect to the set $\{Y_l;l\ge 1\}$.
A direct change of variables provides
\bea
\int_0^{\infty}\!\dd Y_1\ldots \dd Y_n\,
h\left(\frac{2\pi Y_1}{\sum_{k=1}^n Y_k},\ldots,
\frac{2\pi Y_n}{\sum_{k=1}^n Y_k}\right)\exp(-Y_1-\ldots-Y_n)&& 
\nonumber\\[2mm]
=\frac{(n-1)!}{ (2\pi)^{n-1} }\int_0^{2\pi}\dd\eta_1\ldots
\dd\eta_{n}\, h(\eta_1,\ldots,\eta_{n})\,
\Theta\left(\sum_{k=1}^{n}\eta_k-2\pi\right),&&
\eea
which shows that the average with respect to the exponentially distributed
i.i.d.~$Y_k$ is the same as the average weighted with ${\cal P}^{(0)}$.
The same method applies to the $X_m$ and $\xi_m$. 
We now have the following consequence. 
Let us consider the limit $n\to\infty$
with $k$ a fixed integer and $(m_1,m_2,\ldots,m_k)$ 
a not necessarily fixed subset of $\{1,2,\ldots,n\}$.
Then in this limit the marginal
probability distribution of the 
vector $(n/2\pi)(\eta_{m_1},\ldots,\eta_{m_k})$
converges to the probability distribution of 
$(Y_{m_1},\ldots,Y_{m_k})$, {\it i.e.,} to a product of $k$ exponentials. 
An analogous statement holds for the $\xi_m$ and $X_m$.

Let us define the scaled
zero-average variables $\delta x_m$ and $\delta y_m$ by
\bea
n^{-1}\delta x_m &=& \xi_m - 2\pi n^{-1}, \nonumber\\[2mm]
n^{-1}\delta y_m &=& \eta_m - 2\pi n^{-1},
\qquad m=1,\ldots,n.
\label{defdeltaxy}
\eea
They satisfy the sum rules $\summ\delta x_m=\summ\delta y_m=0$
and one readily calculates their variances and covariances,
\bea
\la \delta x_\ell \delta x_{m}\ra_0 &=& \frac{(2\pi)^2 n}{2n+1} 
\Big( \delta_{\ell m}-\frac{1}{n} \Big),
\nonumber \\[2mm]
\la \delta y_\ell \delta y_{m}\ra_0 &=& \frac{(2\pi)^2 n}{n+1}
\Big( \delta_{\ell m}-\frac{1}{n} \Big), 
\nonumber\\[2mm]
\la \delta x_\ell \delta y_{m}\ra_0 &=& 0, \qquad \qquad
\ell,m=1,\ldots,n,
\label{exprcorr}
\eea
which will be needed later. The necessary calculations may be 
carried out directly via (\ref{defP0})
or with the use of the explicit realization of $\xi_m$ and $\eta_m$ as
functions of $X_1,\cdots,X_n$ and $Y_1,\cdots,Y_n$.
Equations (\ref{exprcorr}) show that $\delta x_m$ 
and $\delta y_m$ are of order $n^0$ in the large $n$ limit.

We define $u_m$ by
\bea
n^{-\frac{1}{2} }u_m &=& \tfrac{1}{2}\big(\beta_m-\gamma_m \big)
\nonumber\\[2mm]
&=& \beta_1 -\sum_{\ell=1}^{m-1}(\xi_\ell-\eta_\ell) -\tfrac{1}{2}\xi_m\,,
\qquad m=1,\ldots,n,
\label{defu}
\eea
where to obtain the second line we used (\ref{inversebgxy}).
It follows from (\ref{defu}) together with the scaling
of the $\delta x_m$ and $\delta y_m$ found in \ref{defdeltaxy}
that in the large-$n$ limit 
the $\beta_m$ and $\gamma_m$, being sums of independent random variables,
have Gaussian distributions of average $\pi/n$ and of width 
$\,\sim\! n^{-1/2}$. The $u_m$ therefore remain of order $n^0$ when $n$ gets
large \cite{footnote3}. 
We define $r_m$ by 
\beq
\rho_m = 1 + n^{-\frac{1}{2}}r_m\,, \qquad m=1,\ldots,n.
\label{defr}
\eeq
>From the scaling of $\beta_m$ and $\gamma_m$ together with
(\ref{ratiorho}) and (\ref{defr}) it follows that the $r_m$ are of order 
$n^{0}$. 
Due to the sum rule on the $\rho_m$ in (\ref{sumrho})
they satisfy $\summ r_m=0$.


\subsection{Second order recursion}
\label{secrecursion}

We are now ready to perform the large-$n$ expansion of 
equation (\ref{ratiorho}),
$\rho_m/\rho_{m-1}=\cos\gamma_m/\cos\beta_m$,
using the scaling of the preceding subsection.  
This yields a recursion relation which
in terms of the scaled variables $r_m$, $u_m$, and $\delta x_m$ takes the form
\beq
r_m-r_{m-1} + \ldots = \tfrac{1}{2}(\beta_m^2-\gamma_m^2)+\ldots
\label{diff0tau}
\eeq
where the dots represent terms of higher order.
Using that $\beta_m+\gamma_m=\xi_m$ and $\beta_m-\gamma_m=2n^{-1/2}u_m$ 
we get from this in the large-$n$ limit
\beq
r_m-r_{m-1} = \frac{2\pi}{n} u_m + \frac{1}{n}{\delta x_m} u_m\,,
\label{diff1tau}
\eeq
where now on both sides we have kept only the lowest order.
Here and throughout the remainder it is understood that all quantities
are $n$-periodic in $m$. 
In (\ref{diff1tau}) the first and the second term on the right hand side 
come from the average and the random part of the angle $\xi_m$, respectively.
Both are of order $n^{-1}$, but there is a difference.
To see this we imagine to sum (\ref{diff1tau})  
on an $m$ interval of order $n^\delta$ for some $\delta>0$.
Since $u_m$ has long-range correlations we may 
consider it as effectively constant over this interval.
Then, because of the $\delta x_m$, 
the second term will become a sum of independent zero-average random
variables and hence its contribution will be of relative order $n^{-\delta/2}$ 
with respect to that coming from the first term.
We may therefore neglect the $\delta x_m u_m$ term in (\ref{diff1tau})
when our purpose is to study the variation of $r_m$ on a scale 
that increases as a power of $n$.
Taking now second order differences we find the recursion 
\beq
r_{m-1}-2r_m+r_{m+1} = 2\pi n^{-\frac{3}{2}}f_m\,,
\label{diff2r}
\eeq
with the right hand member defined by
\beq
f_m = -\tfrac{1}{2}\big( \delta x_{m}-2\delta y_{m}+\delta x_{m+1} \big).
\label{deffm}
\eeq
This quantity satisfies the sum rule $\summ f_m=0$.


\subsection{Second order differential equation}
\label{seccontinuum}

We may pass to the following continuum description.
We define the polar angle $\phi=2\pi m/n$ and the functions 
$r(\phi) = r_m$ and $\dd r/\dd\phi = u(\phi) = u_m$. 
In the large-$n$ limit $r$ becomes a continuous function 
of a continuous variable.
For the second order difference
on the left hand side of (\ref{diff2r}) we get
$r_{m-1}-2r_m+r_{m+1} \to (2\pi/n)^2\dd^2 r/\dd\phi^2$.
On the right hand side we put 
$f_m \to 2\pi n^{-\frac{1}{2}}f(\phi)$.
This converts (\ref{diff2r}) into the second order differential equation
\beq
\frac{\dd^2 r(\phi)}{\dd\phi^2} = f(\phi), \qquad 0\leq\phi\leq 2\pi,
\label{randaccprocess}
\eeq
valid for angle differences $|\phi-\phi'|$
on the scale $n^{-1+\delta}$, that is, large on the scale
of the discrete index $m$. 
In the same way as in the previous subsection we may imagine to sum 
(\ref{randaccprocess}) on an $m$ interval of length $\sim n^\delta$.
In the large-$n$ limit $\sum_{m=m_0}^{m_0+n^\delta}f_m$ 
is a Gaussian variable centered at zero, 
whose correlation with similar variables on neighboring intervals is
small. Hence  
on scales $\gsim n^\delta$ the right hand side of (\ref{randaccprocess})
becomes Gaussian noise. Its autocorrelation function 
follows from definition (\ref{deffm}) and 
the correlations (\ref{exprcorr}), namely
\beq
\la f(\phi)f(\phi') \ra_0 = 
\tfrac{3}{2}\big[ 2\pi\delta(\phi-\phi')\,-\, 1 \big].
\label{exprautocorr}
\eeq
The noise is white except for the extra $-1$ term inside the brackets.
This term leaves all Fourier components of $f$ with nonzero wave\-number
unaffected: they behave as under {\it white\,} Gaussian noise.
Only the zero-\-wave\-number component of $f$ is exceptional. 
By integrating (\ref{exprautocorr}) on $\phi$ and $\phi'$ one finds that
$\int_0^{2\pi}\!\dd\phi\, f(\phi)=0$, in agreement with the sum rule in the
last line of section \ref{secrecursion}.
Now the differential equation (\ref{randaccprocess}) 
shows that instead of having this sum rule on $f$ one may equivalently
impose on $r$ that its derivative be periodic, $r'(0)=r'(2\pi)$. 
The functions $r(\phi)$ that we must deal with
are therefore, in summary, those that satisfy
equation (\ref{randaccprocess}) with boundary conditions
\beq
r(0)=r(2\pi),  \qquad  r'(0)=r'(2\pi)
\label{bcrphi}
\eeq
and that, due to the relation in the last line
of section \ref{secscalingI}, obey moreover the zero-integral constraint 
$\int_0^{2\pi}\!\dd\phi\, r(\phi)=0$.
For this class of functions the noise $f$ is effectively white.

In the statistical physics literature equation (\ref{randaccprocess})
represents what is called the {\it Random Acceleration Process} (RAP).
Some time ago this process appeared \cite{Hilhorst05b} 
(but without being named explicitly) 
in a similar way in the study 
of many-sided Voronoi cells. In section \ref{secGMOR} of this work
we will see that, through the connection that we have discovered here,
known facts about the RAP provide further answers to Sylvester's question.


\subsection{Large-$n$ expansion of $p_n^*$}
\label{secexpansion}

We return to the evaluation of the probability $p_n^*$
given by (\ref{expnV}),
\beq
p^*_n = p_n^{(0)}\la \Theta\ee^{-{\mathbb V}} \ra_0\,, \qquad
p_n^{(0)} = \frac{1}{4\pi^2}\,\frac{(8\pi^2)^n}{(2n)!}\,,
\label{expnVbis}
\eeq
in which ${\mathbb V}$ is given by (\ref{defmathbbV}).
All operations that led to these equations
were exact for any finite $n$. We will now see what happens
if we perform a large-$n$ expansion assuming the scaling of section
\ref{secscalingI}.
In the large-$n$ limit the condition imposed by $\Theta(\xi,\eta)$ 
in (\ref{expnV}) will be satisfied with a probability that goes 
exponentially fast to unity; hence we have $\Theta\simeq 1$ and
\beq
p_n^* \simeq p_n^{(0)}\expmVo\,,
\label{expn2}
\eeq
where the $\simeq$ sign denotes an equality up to
corrections that are exponentially small in $n$.
We consider now the $n\to\infty$ limit of ${\mathbb V}$. 
In (\ref{defmathbbV}) we
have $G'(\xi,\eta;\beta_*)=1+{\cal O}(n^{-1})$ \cite{Hilhorst07}. 
Expanding the other factors for small angles we get
\bea
\ee^{-{\mathbb V}} &=& \exp\Big[ 
- \frac{1}{n}\summ \big( r_m^2-u_m^2 \big)
-2n^{1/2}\maxm r_m + \Big( \maxm r_m \Big)^2 + \ldots\Big]
\nonumber\\[2mm]
&\simeq& \exp\Big[ 
- \frac{1}{2\pi}\int_0^{2\pi} \!\dd\phi\,
\Big( r^2(\phi)-\big( \frac{\dd r}{\dd\phi} \big)^2 \Big)
-2n^{1/2}\maxphi r(\phi) 
\nonumber\\[2mm]
&& \phantom{\exp\Big[} + \Big( \maxphi r(\phi) \Big)^2 + \ldots\Big],
\label{exprV}
\eea
where the dots indicate terms of higher order in the angles.
As $n\to\infty$ with the scaling discussed in section \ref{secscalingI}, 
the dot terms in the exponential in (\ref{exprV}) tend to zero;
the first term (which is a sum of $n$ terms compensated by a factor $n^{-1}$)
and the third term stay of order $n^0$;
and the second term is of order $n^{1/2}$.


\subsection{Scaling in the large-$n$ limit. II}
\label{secscalingII}

As shown by (\ref{expn2}),
expression (\ref{exprV}) has to be submitted to an average,
denoted $\la\ldots\ra_0$, with respect to
all solutions $r(\phi)$ of the random acceleration process
satisfying the constraints stated in section \ref{seccontinuum}.
The law for $r(\phi)$ follows, {\it via\,} equation (\ref{randaccprocess}),
from the law for $f(\phi)$, which in turn follows, {\it via\,} (\ref{deffm}),
from the law (\ref{defP0}) for the $\xi_m$ and $\eta_\ell$.
When $n$ is large, the second term in the
exponential in (\ref{exprV}) will suppress large excursions of
$r(\phi)$ from zero in a severe but as yet quantitatively unknown way.
Let us therefore suppose that the average $\expmVo$ 
draws its principal contribution from those $r(\phi)$
that stay within a tube
of a width $w_n \sim n^{-\alpha}$ with an as yet unknown $\alpha$. 
Hence the effectively contributing $r(\phi)$ will scale as
\beq
r(\phi) \sim n^{-\alpha}, \qquad \maxphi r(\phi) \sim n^{-\alpha}
\label{newscaling}
\eeq
for some as yet unknown $\alpha>0$.
We must ask ourselves, first of all, if we are allowed to
have the scaling already introduced above followed by this new scaling.
In fact we are, because it only restricts us more
narrowly to the center of the realm where the first
scaling holds. The same results are arrived at if the full combined 
scaling is used from the start.
The reason for the procedure we chose 
is that this is the best way of showing the
connection with the Random Acceleration Process, of which
finally only a subprocess plays a role, namely
the one consisting of trajectories that stay within a tube.

Secondly, if indeed the main contributing $r(\phi)$ 
scale with a negative power of $n$, then the first and the third term
in (\ref{exprV}) are negligible with respect to the second one
in a large-$n$ expansion.
So finally the problem becomes to calculate the average
$\la\ee^{-{\mathbb V}}\ra_0$ with 
\beq
\la\ee^{-{\mathbb V}}\ra_0 
\simeq \big\la \exp\big[ -2n^{1/2}\max_{0\leq\phi\leq 2\pi}r(\phi) 
\big]\big\ra_0\,,
\label{exprPrmax}
\eeq
where in the last line the average is with respect to the  
process $r(\phi)$ defined by (\ref{randaccprocess}) and (\ref{exprautocorr}).
In equation (\ref{exprPrmax}) the only reference to $n$ is the one explicitly
visible in the exponential on the right hand side.
Hence here $n$ has become an ``external'' parameter 
coupling to the maximum value 
of an $n$ independent random acceleration process. 
Finally, the remainder ${\cal R}_n$ in the expansion of the convexity
probability $p_n(D)$ in a disk is related to (\ref{exprPrmax}) by
\beq
{\cal R}_n = \log\expmVo.
\label{rel}
\eeq
The connection, in the large-$n$ limit, 
between Sylvester's question and the Random Acceleration Process  
is the principal achievement of this paper.
We will now see that, when
combined with what is known about this process, it leads to the results
announced in the introduction.


\section{Properties of the Random Acceleration Process}
\label{secproperties}


\subsection{The work by Gy\"orgyi {\it et al.}}
\label{secGMOR}

The random acceleration process is the $k=2$
member of the wider class of equations
\beq
\dd^k\tilde{r}/\dd\phi^k=\tilde{f}(\phi)
\label{kRAP}
\eeq
with $k=1,2,\ldots$, and where $\tilde{f}(\phi)$ is white
Gaussian noise.
Because of their relevance in physics and in applied statistics,
many examples of such processes have been studied.

In early work on the $k=1$ case, where $\tilde{r}(\phi)$ is Brownian motion,
Foltin {\it et al.} \cite{FORWZ94} 
analyzed the distribution of the square width of $\tilde{r}(\phi)$.
Burkhardt \cite{Burkhardt93} considered the $k=2$ problem in a half-space. 
More recently, Majumdar and Comtet \cite{MajumdarComtet04,MajumdarComtet05},
were interested in the distribution of the ``maximum height''
$r_{\rm max}$, defined as the
maximum of $\tilde{r}(\phi)$ relative to its interval average,
\beq
r_{\rm max} \equiv \maxphi\tilde{r}(\phi) 
-\frac{1}{2\pi}\int_0^{2\pi}\!\dd\phi\,\tilde{r}(\phi).
\label{defrmax}
\eeq
Using powerful path integral methods, 
Majumdar and Comtet showed that for $k=1$ and periodic boundary conditions
the probability law of the maximum height is the Airy distribution.
They pointed out the importance of this problem as an instance of extreme value
statistics of a set of strongly {\it correlated\,} random variables.
The equation with general $k$ was studied very recently
by Gy\"orgyi {\it et al.} \cite{GMOR07} (GMOR), who also focused
on the distribution $P(r_{\rm max})$ of the maximum height (\ref{defrmax}). 
The authors of reference \cite{GMOR07} calculate this
distribution for the class of functions $\tilde{r}(\phi)$
that satisfy (for $k=2$) the periodicity conditions
\beq
\tilde{r}(0)=\tilde{r}(2\pi), \qquad \tilde{r}'(0)=\tilde{r}'(2\pi).
\label{periodicity}
\eeq
Hence, after subtraction of the interval average, 
this is the same class 
that appears in the averages $\la\ldots\ra_0$ in sections 
\ref{secexpansion} and \ref{secscalingII}.
This establishes the applicability of GMOR's results to the problem
of section \ref{secracp}.

By extending Majumdar and Comtet's \cite{MajumdarComtet04} methods GMOR
found an expression for the distribution $P(r_{\rm max})$ of $r_{\rm max}$.
Still more recently, Burkhardt {\it et al.} considered the distribution 
of the maximum relative to the initial value \cite{Burkhardtetal07}. 

Expression (\ref{exprPrmax}) is
the Laplace transform with Laplace variable $2n^{1/2}$ 
of the maximum height distribution
of the random function $r(\phi)$. No mathematically rigorous results 
for this quantity seem to exist; however,
closely related properties, involving in
particular the maximum of the {\it absolute value\,} $|r(\phi)|$, were
studied, {\it e.g.,} by Khoshnevisan and Shi \cite{KhoshnevisanShi98}
and more recently by Chen and Li \cite{ChenLi03}.
The Laplace transformed maximum height distribution did figure
among the many properties of the RAP studied by
GMOR \cite{GMOR07}. These authors express this quantity
by means of a trace formula 
involving the second order differential operator
\beq
{\cal L} = \frac{1}{2}\frac{\partial^2}{\partial u^2}
 - u\frac{\partial}{\partial r} - r,
\label{defL}
\eeq
which is the generator of
the Markov process $\big( r(\phi),u(\phi) \big)$.
They consider the linear eigenvalue problem 
\begin{align}
{\cal L}\psi(r,u)&=-\epsilon\psi(r,u),
\qquad & &0<r<\infty, \quad -\infty<u<\infty, \nonumber \\[2mm]
\psi(0,u)&=0, \qquad &  &u>0,
\label{evproblem}
\end{align}
and assume that it has a well-defined solution with lowest eigenvalue
$\epsilon_0$. 
When converted to our notation \cite{footnote2}, the result of GMOR,
obtained essentially by a scaling of all variables in the trace formula,
is that in the limit of large $n$
\beq
-\log \expmVo \simeq 2\epsilon_0(3\pi^4 n)^{\frac{1}{5}},
\qquad n\to\infty.
\label{resultGmor}
\eeq
With (\ref{resultGmor}) we have obtained our main result, 
equation (\ref{mainresult}) of section \ref{secintroduction}.


\subsection{Heuristic argument for the power $1/5$}
\label{secheuristic}

\begin{figure}
\begin{center}
\scalebox{.45}
{\includegraphics{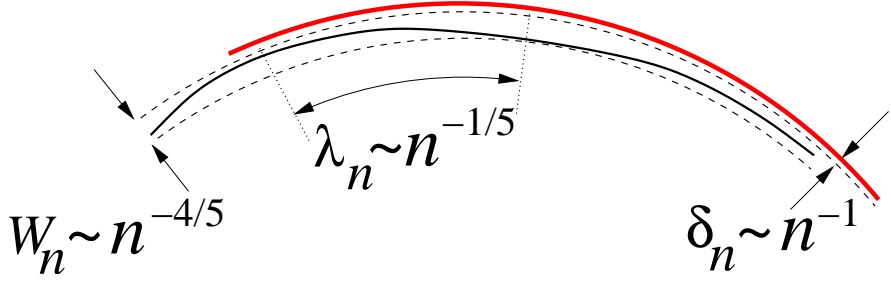}}
\end{center}
\caption{\small
Heavy solid arc: edge of the disk of radius $R_D=1$.
Solid curve: the convex polygon $R(\phi)$ in the limit $n\to\infty$. 
Dashed arcs: annular tube of width $W_n\sim n^{-4/5}$.
The two dotted transverse line segments mark the ends of a
typical tube section of length $\lambda_n$  
between two consecutive 
`reflections' of the polygon at the inner and the outer tube walls; 
the section length scales as $\lambda_n \sim n^{-1/5}$. 
The distance of closest approach 
of the edge scales as $\delta_n\sim n^{-1}$.} 
\label{figtube} 
\end{figure}

We will derive the power
$\frac{1}{5}$ found by GMOR by a heuristic scaling argument.
This is not meant as additional evidence for this result,
but as a physical way of understanding it. 
We focus on the $n$ dependence of
$\expmVo$. If we agree to consider the expression for ${\mathbb V}$
in the exponential in (\ref{exprPrmax}) as an `energy', 
then what we will present is basically an energy-versus-entropy argument.

We begin by supposing that due to
the factor $n^{1/2}$ included in the `energy'
in (\ref{exprPrmax}) the contributions to
$\expmVo$ will come essentially from trajectories $r(\phi)$ 
with a maximum less than some small but as yet unknown width $w_n$. 
These contributing trajectories 
will typically be confined to a tube of width $2w_n$. 

We refer now to figure \ref{figtube}.
We imagine the tube divided into sections 
such that inside each section the trajectory evolves freely,
but at the sections' ends it is `reflected' by the tube walls.
Let $\lambda_n$ be the angular interval of a typical section (and hence
$\lambda_n\Rav$ is its length).
We may estimate the $n$ dependence of $\lambda_n$ and $w_n$ as follows.
If $u(\phi) = \dd r/\dd\phi$ is the `radial speed',
then (\ref{randaccprocess})
may be rewritten $\dd u/\dd\phi = f(\phi)$. 
In an angular interval $\lambda_n$ this speed will change by
$\delta u = \int_{\phi}^{\phi+\lambda_n}\dd\phi' f(\phi') \sim 
\lambda_n^{1/2}$, so that the typical radial speed 
inside the tube will be of order $|\delta u| \sim \lambda_n^{1/2}$.
It follows that in this interval
the value $r(\phi)$ of the radius itself will change by
$\delta r \sim \lambda_n \delta u \sim \lambda_n^{3/2}$.
This means that we have typically
$|r|\sim \maxphi r(\phi) \sim w_n \sim \lambda_n^{3/2}$.
Hence a typical contributing trajectory $r(\phi)$ will contribute to
the average on the right hand side of (\ref{exprPrmax})
an amount $\exp(-{\rm cst}\times n^{1/2}\times \lambda_n^{3/2})$.
This is our result for the `energy' factor.

We now estimate which fraction of all 
trajectories is `contributing' in the above sense,
that is, stays within the tube of width $2w_n$. 
This confinement costs a constant amount of entropy per section, due to the
`reflection' at its ends, the trajectory being essentially free everywhere 
else.
Since the number of sections is $2\pi/\lambda_n$, our result for the `entropy'
factor is $\exp(-{\rm cst}\times 2\pi\lambda_n^{-1})$. 

Upon multiplying the energy and entropy estimates we
arrive at the heuristic estimate for $\expmVo$. Taking logarithms yields
\beq
-\log\expmVo \,\simeq\, C_1\lambda_n^{-1}+C_2 n^{1/2}\lambda_n^{3/2},
\qquad n\to\infty,
\label{expmVoheur}
\eeq
where $C_1, C_2>0$. In (\ref{expmVoheur})
the $n$ dependence of the section length $\lambda_n$ is still
unspecified. By setting $\lambda_n \sim n^{-\alpha}$
and minimizing the right hand side with respect to
$\alpha$ we find $\alpha=\frac{1}{5}$, whence
\beq
\lambda_n \sim n^{-1/5}, \qquad w_n \sim n^{-3/10}.
\label{reselln}
\eeq
We may return now to the original unscaled variables. By transposing 
(\ref{defRavrm}) to continuum notation we see that the radius
of the polygon, $R(\phi)=\Rav[1+n^{-1/2}r(\phi)]$,
runs through a tube of width $W_n = \Rav n^{-1/2}w_n \sim n^{-4/5}$,
as depicted in figure \ref{figtube}.


\subsection{Exact bounds}
\label{secweaker}

The mathematical methods used in section \ref{secdisk} are fully exact.
Those of section \ref{secracp} are basically a systematic expansion in
negative powers of $n$; although we did not rigorously prove its validity, it
has been obtained by methods that commonly lead to exact results.
The argument of section \ref{secGMOR}, due to GMOR, involves an assumption
of a different kind, namely the existence of a lowest-energy solution of 
(\ref{defL}) and (\ref{evproblem}) with an eigenvalue $\epsilon_0$.
We have no reasons to doubt that this solution exists.
However, we wish to point out here that without this existence assumption
one may derive by elementary methods a weaker but still very meaningful
statement. We have relegated the proof to Appendix A.
We there show by straightforward methods 
starting from (\ref{exprPrmax}) that we have the bounds 
\beq
c_6 n^{\frac{1}{6}} \leq -\log\expmVo \leq c_4 n^{\frac{1}{4}},  \qquad
n\to\infty,
\label{exprbounds}
\eeq
with positive constants $c_4$ and $c_6$.

Hence, with or without the existence assumption of $\epsilon_0$,
it is established in either case that the large-$n$ expansion of
$\,\log\expmVo\,$ depends nonanalytically on $n$.
Consequently, because of (\ref{rel}) and (\ref{asptD}),
the expansion of $\log p_n(D)$ for a disk contains this same nonanalytic term
and is clearly distinct from the expansions (\ref{asptST}) for 
squares and triangles.


\section{Closest approach of the edge}
\label{secapproach}

Let us call $\delta_n$ the distance of closest approach between the perimeter
of the polygon and the edge of the unit disk, that is,
\beq
\delta_n = 1-\maxm R_m.
\label{defdeltan}
\eeq
Given the results of the preceding sections, it is now fairly easy
to study this quantity. 

We now first turn to the study of the distribution
of $\Rav$, which we will call ${\sf P}(\Rav)$.
We have
\beq 
{\sf P}({\Rav}) = \frac{{p}({\Rav})}{p_n^*},
\label{defsfp}
\eeq
where the denominator $p_n^*$ is given by integral (\ref{expn})
and the numerator ${p}({\Rav})$ by this same integral
except that we put primes on the variables of integration 
and insert a factor $\delta(\Rav'-{\Rav})$.
Evaluation of the numerator proceeds as it did for the
denominator in section \ref{sectransfI}. Instead of 
(\ref{intonRav}) we have to calculate
\beq
\int_0^{\minm\rho_m^{-1}}\!\dd\Rav'\,\Rav'^{2n-1}\,
\delta\big( \Rav'-{\Rav} \big) \,=\,
\Rav^{2n-1}\theta
\big( \big[ \maxm\rho_m \big]^{-1} - {\Rav} \big).
\label{intonRavbis}
\eeq
It then follows that 
\beq
{\sf P}(\Rav) = \frac{\la \Theta\ee^{ -{\mathbb V}_{\!\Rav} }\ra_0}
{\la \Theta\ee^{ -{\mathbb V} }\ra_0}\,,
\label{exprsfp}
\eeq
in which $\exp(-{\mathbb V}_{\!\Rav})$ is the same expression as 
(\ref{defmathbbV}) except that the factor 
$[ \maxm\rho_m ]^{-2n}$
has been replaced with
$2n \Rav^{2n-1} \big( [ \maxm\rho_m ]^{-1} - {\Rav} \big)$.
Expression (\ref{exprsfp}) for ${\sf P}(\Rav)$ is still fully exact.

We cannot go beyond this point unless we make again a large-$n$ expansion.
We define a variable $\dav$ by
\bea
\Rav = 1- n^{-1/2}\dav,
\label{defdav}
\eea
where we tacitly suppose that $\dav$ is at most of order  $n^0$ but
possibly smaller. 
Using in (\ref{defdeltan}) that $R_m=\Rav\rho_m$, 
employing (\ref{defdav}), 
and expanding, we obtain between $\delta_n$ and $\dav$  
the relationship
\beq
\delta_n = n^{-1/2}\big[ \dav - \maxm r_m \big]\,+\,\ldots,
\label{exprdeltan}
\eeq
where the dots indicate terms of higher order.

Using expression (\ref{defr}) for $\rho_m$ and expanding as
in section \ref{secexpansion} we get an expression for ${\sf P}(\Rav)$
in terms of the variable $\dav$,
\beq
{\sf P}(\Rav) = 2n\,\la\ee^{-{\mathbb V}}\ra_0^{-1}
\Big\la\!\exp\Big[ -\frac{1}{n}\summ \big(r_m^2-u_m^2\big) 
-2n^{1/2}\dav - \dav^2 +\ldots \Big]\Big\ra_0.
\label{exprsfp2}
\eeq
For the same reasons as in section \ref{secexpansion}
we neglect now the sum and the $\dav^2$ term in the exponential
in (\ref{exprsfp2}).
Finally, using (\ref{exprdeltan}), we eliminate the variable $\dav$
from (\ref{exprsfp}) in favor of $\delta_n$ whose
distribution we will call ${\sf p}(\delta_n)$.
We get in the limit $n\to\infty$ the final result of this section,
namely the probability distribution {\sf p} of 
the distance of closest approach of the edge,
\beq
{\sf p}(\delta_n) \,=\, 2n\,\ee^{-2n\delta_n}, \qquad \delta_n>0.
\label{exprqdeltan}
\eeq
Not only does this show that $\delta_n$ 
scales as $1/n$, but also that it is exponentially distributed.


\section{Conclusion}
\label{secconclusion}

In this work we considered Sylvester's question: given $n$ points chosen
randomly from a uniform distribution on a disk, what is the probability that
they are the vertices of a convex $n$-sided polygon? For $n$ large this
probability becomes very small and one may consider the asymptotic expansion
of $\log p_n$. The first two terms of the expansion were known; they are
proportional to $n\log n$ and to $n$.
In this work we establish, first of all, 
a relation between Sylvester's question
and the Random Acceleration Process.
We then show that the third term in the expansion (in absolute value)
is asymptotically bounded from above and below by
$\sim n^{1/6}$ and $\sim n^{1/4}$, respectively, so that it
must be nonanalytic in $n$.
If one accepts the hypothesis underlying the work of Gy\"orgyi {\it et al.},
which we easily do, it follows that this term must be proportional to
$\sim n^{1/5}$ with a well-defined coefficient given in (\ref{mainresult}).
Along with this expansion of $p_n$ we harvest 
a variety of results concerning the most
probable way that the $n$ points are distributed along the edge of the disk.

The subject has not been fully exhausted. Remaining questions 
in the present work concern, for
example, correlations between the angles. Besides, a natural generalization would be to consider more general convex domains $K$ and in particular convex polygons. In this context, the work \cite{Buchta06} provides precise results on convex chains for deriving exact distributional results on the number of vertices of the convex envelope. Extensions of this work 
could deal with the crossover phenomenon that must exist when the convex
domain $K$ varies between finite-sided polygons and the disk: for example,
what happens when one considers
$n$ points randomly chosen 
in the interior of a regular $M_n$-gon when $n$ and $M_n$ 
tend to infinity in a specified way?
Another extension would be the study of the convex envelope
of $N_n$ randomly chosen points knowing that this envelope
is $n$-sided, again in the limit of $n$ and $N_n$ tending to infinity.
However, we leave these and other matters for future investigation.


\appendix


\section{Bounds for $\expmVo$\, as\, $n\to\infty$}
\label{secbounds}

In order to derive upper and lower bounds on the average $\expmVo$
we begin by expressing this quantity in terms of the Fourier transforms of the
variables involved.

\subsection{Fourier transforms}
\label{secFT}

We define the Fourier transform 
\beq
\hat{f}_q= \frac{1}{2\pi n^\half}\summ \ee^{2\pi{\rm i} qm/n} f_m\,,
\qquad
\hat{r}_q=n^{-1}\sum_{m=1}^n \ee^{2\pi{\rm i} qm/n} r_m\,,
\label{deffqrq}
\eeq
where, if for convenience we take $n$ odd,
$q=0,\pm 1,\pm 2,\ldots,\pm(\frac{1}{2}n-\frac{1}{2})$.
The sum rules imply that $\hat{f}_0=\hat{r}_0=0$.
In Fourier language recursion (\ref{diff2r}) becomes
\beq
\hat{r}_q = -\frac{\pi^2}{ n^2 \sin^2\frac{\pi q}{n} } \hat{f}_q\
\,\,\simeq\,\, -\frac{1}{q^2}\hat{f}_q\,,
\qquad q \neq 0.
\label{diff2rq}
\eeq
where the $\simeq$ sign indicates the limit $n\to\infty$ at fixed $q$.
This amounts to neglecting higher orders in $q/n$,
which are small if in agreement with our preceding discussion
we restrict ourselves to $q$ on a scale $\sim n^{1-\delta}$,
that is, to spatial distances which in units of the index $m$
scale as $\sim n^\delta$.
>From definition (\ref{deffqrq}) together with (\ref{deffm}) 
we have furthermore
\beq
\la\hat{f}_q\hat{f}_{q'} \ra_0 = \tfrac{3}{2}\delta_{q+q',0}\,,
\qquad q,q'\neq 0.
\label{exprcorrf}
\eeq
The $\hat{f}_q$ are distributed according to a marginal distribution 
$P^{(0)}(f)$ of ${\cal P}^{(0)}(\xi,\eta)$. 
In the limit $n\to\infty$ all $\hat{f}_q$ become Gaussian distributed 
and therefore $P^{(0)}$ is given by
\beq
P^{(0)}(f) = \prod_{q \neq 0}
\frac{1}{\sqrt{3\pi}} \exp\Big(\!-\tfrac{1}{3}\hat{f}_q\hat{f}_{-q} \Big).
\label{exprPf}
\eeq
This distribution may be used for averaging quantities
that depend essentially on the long wavelength properties of the process.

It will be easier to deal with real, as opposed to complex, quantities.
Choosing a convenient normalization
we define the radial and angular components $F_q$ and $\omega_q$ of
$\hat{f}_{\pm q}$ by 
\beq
\hat{f}_{\pm q} = \tfrac{1}{2}\sqrt{6}\, 
F_q (\cos\omega_q \pm {\rm i}\sin\omega_q), \qquad q>0.
\label{deffq}
\eeq
Letting as before $\phi=2\pi m/n$, we have that
in terms of these Fourier transforms the process $r(\phi)$ becomes
\beq
r(\phi) = \sum_q\ee^{-{\rm i}q\phi}\,\hat{r}_q 
= -\sqrt{6}\,\sum_{q=1}^\infty 
\frac{F_q}{q^2}\cos\big(q\phi-\omega_q\big).
\label{exrf}
\eeq
Substituting (\ref{exrf}) in (\ref{exprPrmax})
and transforming to the $F_q$ and $\omega_q$ as new variables of integration
we get
\bea
\la \ee^{-{\mathbb V}} \ra_0 &\simeq& 
\int_0^\infty \prod_{q=1}^\infty 2F_q\dd F_q
\int_0^{2\pi} \prod_{q=1}^\infty \frac{\dd\omega_q}{2\pi}
\exp\Big[ -\sum_{q=1}^\infty F_q^2 \Big]
\nonumber\\[2mm]
&& \times 
\exp\Big[ -2\sqrt{6n} \maxphi\sum_{q=1}^\infty
\frac{F_q}{q^2}\cos\big(q\phi-\omega_q \big) \Big].
\label{exVF}
\eea
In the right hand side of (\ref{exVF}) the only $n$ dependence that comes
in is the one explicitly exhibited in the argument of the exponential.
We note that the Fourier expression (\ref{exrf}) shows that $r(\phi)$,
due to the factor $1/q^2$ in the sum on $q$,
is precisely a quantity of the type that receives its main contribution from
small $q$ values and whose properties may therefore be calculated 
by averaging with respect to (\ref{exprPf}). 
The feature that makes this problem hard to solve exactly
is the maximum that is required in (\ref{exVF}). 
In what follows we will obtain upper and lower bounds for $\expmVo$.


\subsection{Two inequalities}
\label{secineq}

We recall that the argument of the second exponential in (\ref{exVF}) is
equal to $-2\sqrt{n}\maxphi r(\phi)$.
Upper and lower bounds for $\expmVo$ are based on bounds
for this maximum expressed in terms of the Fourier amplitudes of $r(\phi)$.
These bounds, valid asymptotically for $n$ large, read
\beq
\frac{\sigma^2}{M}
\leq \maxphi {r}(\phi) \leq M,
\label{boundsRPhi}
\eeq
where
\beq
M = \sqrt{6} \sumq \frac{F_q}{q^2}\,,
\qquad
\sigma^2 = 3 \sumq \frac{F_q^2}{q^4}\,,
\label{defMsigma}
\eeq
the $F_q$ are defined by (\ref{exrf}),
and in which it is assumed that both sums on $q$ converge. 
The upper bound given by (\ref{boundsRPhi})-(\ref{defMsigma}) is obvious
and leads to a lower bound for $\expmVo$
derived in Appendix \ref{seclowerbound}.
The lower bound given by (\ref{boundsRPhi})-(\ref{defMsigma})
is proved in Appendix \ref{seclowerR}; it leads to an upper bound for
$\expmVo$ which is derived in Appendix \ref{secproofupperbound}.


\subsection{Lower bound for $\expmVo$}
\label{seclowerbound}

In this section of the appendix we abbreviate
\bea
N=2\sqrt{6n}.
\label{defN}
\eea
We will find a lower bound for $\expmVo$ valid in the limit of large $N$.
We replace the maximum in (\ref{exVF}) by its upper bound
given in (\ref{boundsRPhi}).
Since this bound is independent of the $\omega_q$, 
the integrals on these variables
reduce to a factor unity. The integrals on the $F_q$ factorize and we get
\beq
\expmVo \geq \prodq\, \int_0^\infty 2x\dd x \exp\Big( -x^2 -Nq^{-2}x
\Big) 
= \exp\Big[ -\sumq {\cal F}(qN^{-\frac{1}{2}}) \Big] 
\label{exprInF}
\eeq
where 
\beq
{\cal F}(\kappa) = - \log \int_0^\infty 2x\dd x\, \exp( -x^2 -\kappa^{-2}x ).
\label{defF}
\eeq
This function is positive for all $\kappa>0$ and it is integrable since
it behaves as ${\cal F}(\kappa)\simeq 4\log \kappa^{-1}$ for $\kappa\to 0$
and as ${\cal F}(\kappa)\simeq\frac{1}{2}\sqrt{\pi}\kappa^{-2}$ 
pour $\kappa\to\infty$.
The sum on $q$ in (\ref{exprInF}) makes the argument of ${\cal F}$ vary by
steps of spacing $N^{-\frac{1}{2}}$ and we may therefore in the limit
$N\to\infty$ replace it by an integral, which gives
\bea
\expmVo &\geq& 
\exp\Big[ -N^{\half}\int_0^\infty\!\dd\kappa\,{\cal F}(\kappa) \Big]
\nonumber\\[2mm]
&=& \exp\big[-c_4n^{ \frac{1}{4} } \big],
\label{exprInintkappa}
\eea
where in the last step we used the relation between $N$ and $n$ given
in (\ref{defN}), and where $c_4$ is a positive constant.


\subsection{Upper bound for $\expmVo$}
\label{secupperbound}

In order to prove the upper bound for $\expmVo$, we first need to prove the
lower bound on $r(\phi)$ stated in (\ref{boundsRPhi}).


\subsubsection{Lower bound for $\maxphi r(\phi)$}
\label{seclowerR}

We prove the following property.\\ 
\noindent {\it Property.\,\,}
Let $f(x)$ be a 
function on $[0,2\pi]$ such that
\bea
\int_0^{2\pi}\dd x f(x) &=& 0,
\label{one}\\[2mm]
\int_0^{2\pi}\dd x f^2(x) &=& 2\pi\sigma^2,
\label{two}\\[2mm]
\min_{0\leq x\leq 2\pi} f(x) &=& -M,
\label{three}
\eea
where $\sigma$ and $M$ are given positive constants. Then
\beq
\max_{0\leq x\leq 2\pi} f(x) \geq 
\tfrac{1}{2}M \,\Big[ \Big( 1+\frac{4\sigma^2}{M^2} \Big)^{1/2}\,-\,1\, \Big].
\label{property}
\eeq
In the limit $M\gg\sigma$ this yields
\beq
\max_{0\leq x\leq 2\pi} f(x) \geq \frac{\sigma^2}{M}
\,\Big[ 1+{\cal O}\Big(\frac{\sigma^2}{M^2}\Big) \Big].
\label{limproperty}
\eeq
{\it Proof.}
We set
\beq
f(x)=f_+(x) - f_-(x)
\label{decompf}
\eeq
in which $f_\pm(x)=\max\{0,\pm f(x)\}$. 
In terms of these Eqs.\,(\ref{one})-(\ref{three}) become
\bea
\int_0^{2\pi}\dd x f_+(x) - \int_0^{2\pi}\dd x f_-(x) &=& 0,
\label{oneprime}\\[2mm]
\int_0^{2\pi}\dd x f_+^2(x) + \int_0^{2\pi}\dd x f_-^2(x) &=& 2\pi\sigma^2,
\label{twoprime}\\[2mm] 
\max_{0\leq x\leq 2\pi}f_-(x) &=& M.
\label{threeprime}
\eea
We abbreviate
\beq
m=\max_{0\leq x\leq 2\pi} f_+(x).
\label{defm}
\eeq
We now have the estimate
\bea
m &\geq& \frac{1}{2\pi}\int_0^{2\pi}\dd x f_+(x)\nonumber\\[2mm]
     &=& \frac{1}{2\pi}\int_0^{2\pi}\dd x f_-(x),
\label{est1}
\eea
where in the second step we used (\ref{oneprime}).
Next,
\bea
\int_0^{2\pi}\dd x f_-(x) &\geq& \frac{1}{M}\int_0^{2\pi}\dd x f_-^2(x)
\nonumber\\[2mm]
&=& \frac{1}{M}\big[ 2\pi\sigma^2 - \int_0^{2\pi}\dd x f_+^2(x) \big]
\nonumber\\[2mm]
&\geq& \frac{2\pi}{M}\big[ \sigma^2-m^2\big],
\label{est2}
\eea
where in the first step we used (\ref{threeprime}),
in the second step (\ref{twoprime}),
and in the third step (\ref{defm}).

We now combine (\ref{est1}) and (\ref{est2}) to get
\beq
m\geq\frac{\sigma^2-m^2}{M}.
\label{estfin}
\eeq
By solving the associated second-order equation for $m$ and using the
definition (\ref{defm}) we are led directly to (\ref{property}).
This completes the proof of property (\ref{property}).

{\it Application.}
We apply inequality (\ref{property}) 
to the function ${r}(\phi)$ given in (\ref{exrf}). 
Its $M$ satisfies $M<M_0$ with $M_0$ given in equation (\ref{defMsigma}) 
and its $\sigma^2$ is given in that same equation.
Since (as is implied by the arguments of section \ref{secproofupperbound}) 
for $n\to\infty$  the ratio $\sigma/M_0$ tends to zero,
we may replace (\ref{property}) by (\ref{limproperty}) and obtain
the desired lower bound stated in (\ref{boundsRPhi}).


\subsubsection{Proof  of  upper  bound  for  $\expmVo$}
\label{secproofupperbound}

We will next find an upper bound for $\expmVo$ valid in the limit of large $n$.
We replace the maximum 
in (\ref{exVF}) by its lower bound given in (\ref{boundsRPhi}).
The integrals on the variables $\omega_q$
reduce again to a factor unity. The integrals on the $x_q$ 
do not factorize in this case. In order to make them do so we introduce
integral representations and write
\bea
\expmVo &<& \int_0^\infty \prodq 2F_q\dd F_q \exp\Big[
-\sumq F_q^2-\tfrac{1}{2}N\sumq\frac{F_q^2}{q^4} \Big/ 
\sumq\frac{F_q}{q^2} \Big]
\nonumber\\[2mm]
&=& \int_0^\infty\dd\mu \int_{-\infty}^\infty\frac{\dd s}{2\pi} 
\int_0^\infty\prodq 2F_q\dd F_q
\exp\Big[ {\rm i}s\Big( \sumq\frac{F_q}{q^2} -\mu \Big)
-\sumq F_q^2 -\frac{N}{2\mu} \sumq\frac{F_q^2}{q^4}
\Big]
\nonumber\\[2mm]
&=& \int_0^\infty\dd\mu\, \int_{-\infty}^\infty\frac{\dd s}{2\pi}\, 
\ee^{{\rm i}s\mu}\,
\exp\Big[ \sumq\log\int_0^\infty 2x\dd x
\Big(\! -x^2-\frac{N}{2\mu q^4}x^2 + \frac{{\rm i}s}{q^2}x \Big)
\Big] 
\nonumber\\[2mm]
&=& \int_0^\infty\dd\mu 
\exp\Big[ -{\cal G}_1 \Big(\frac{N}{2\mu}\Big) \Big]
\int_{-\infty}^\infty\frac{\dd s}{2\pi}\, 
\ee^{{\rm i}s\mu}\,
\exp\Big[ -{\cal G}_2 \Big(s,\frac{N}{2\mu}\Big) \Big]
\label{exprInG}
\eea
in which
\bea
{\cal G}_1(z) &=& 2\sumq\log\alpha_q(z), 
\qquad
\alpha_q^2(z) = 1 + \frac{z}{q^4}\,,
\label{defG1}
\nonumber\\[2mm]
{\cal G}_2(s,z) &=& -\sumq\log\int_0^\infty 2y\dd y
\exp\Big(\! -y^2 + \frac{{\rm i}s}{q^2\alpha_q(z)}y \Big).
\label{defG2}
\eea
We will need the large $z$ behavior of ${\cal G}_1(z)$. 
Upon scaling
$q=\kappa z^{\frac{1}{4}}$  in (\ref{defG1}) we find 
\beq
{\cal G}_1(z)\simeq a_0 z^{\frac{1}{4}}, \qquad
a_0=\int_0^\infty\!\dd\kappa \log(1+\kappa^{-4}), \qquad z\to\infty.
\label{calG1aspt}
\eeq
In order to study ${\cal G}_2(s,z)$ we expand the integrand in its definition
(\ref{defG2}) in a power series in ${\rm i}s$, do the $y$ integration term by
term, and expand the logarithm in powers of ${\rm i}s$. This gives
\bea
{\cal G}_2(s,z) &=& -\sum_{q=1}^\infty \log \Big[ 
\sum_{k=0}^\infty \frac{\Gamma(\tfrac{1}{2}k+1)}{k!} 
\Big( \frac{{\rm i}s}{q^2\alpha_q} \Big)^k \Big]
\nonumber\\[2mm]
&=& -\pi^{\frac{1}{2}}A_1(z){\rm i}s - \tfrac{1}{2}(1+\pi)A_2(z)({\rm i}s)^2 
+\ldots
\label{calG2ser}
\eea
where
\beq
A_k(z)=\sum_{q=1}^\infty\frac{1}{[q^2\alpha_q(z)]^k},
\qquad k=1,2,\ldots\,.
\label{defAk}
\eeq
After again scaling $q=\kappa z^{\frac{1}{4}}$ 
we get by the same method as above
\beq
A_k(z) = a_k z^{\frac{1}{4}-\frac{1}{2}k}\big[\,1\,+ \dots\big], \quad
a_k=\int_0^\infty\! \dd\kappa (1+\kappa^4)^{-\frac{1}{2}k}, \quad 
z\to\infty,
\label{exprAkaspt}
\eeq
where the dots stand for a series in powers of $z^{-\frac{1}{2}}$.
Let us denote by $J_n\big(\mu,\frac{N}{2\mu}\big)$ the integral on $s$
in the last line of (\ref{exprInG}).  Substitution of the above expansion of
${\cal G}_2$ yields
\bea
J_n\big(\mu,\tfrac{N}{2\mu}\big) &=&
\int_0^\infty \frac{\dd s}{2\pi}
\exp\Big[ -{\rm i}s\Big\{ \mu - \pi^{\frac{1}{2}}a_1
\Big( \tfrac{N}{2\mu} \Big)^{-\frac{1}{4}} +\ldots \Big\}
\nonumber\\[2mm]
&& 
-\tfrac{1}{2}s^2(1+\pi)\Big\{ a_2\Big(\tfrac{N}{2\mu}\Big)^{-\frac{3}{4}} 
+\ldots \Big\} 
+ {\cal O}\Big(\Big(\tfrac{N}{2\mu}\Big)^{-\frac{5}{4}}s^3\Big) \Big],
\nonumber\\
\label{exprJn1}
\eea
in which the dots come from expansion (\ref{exprAkaspt}).
Let us scale $s=\big( \frac{N}{2\mu} \big)^{\frac{3}{8}}t$.
We then get
\bea
J_n\big(\mu,\tfrac{N}{2\mu}\big) &=&
\Big(\frac{N}{2\mu}\Big)^{\frac{3}{8}}
\int_0^\infty \frac{\dd t}{2\pi}
\nonumber\\[2mm]
&& \times\,
\exp\Big[ -{\rm i}\Big\{ \mu\Big(\frac{N}{2\mu} \Big)^{\frac{3}{8}}
- \pi^{\frac{1}{2}}a_1\Big( \frac{N}{2\mu} \Big)^{\frac{1}{8}} \Big\}t 
- \tfrac{1}{2}(1+\pi)a_2t^2\,\Big]+\ldots
\nonumber\\[2mm]
&=& 
\Bigg[ \frac{2\pi}{(1+\pi)a_2} \Bigg]^{\frac{1}{2}}
\Big( \frac{N}{2\mu} \Big)^\frac{3}{8}
\exp\Biggl[ - \dfrac
{ 
\Big\{ \mu\Big(\dfrac{N}{2\mu} \Big)^{\frac{3}{8}} 
- \pi^{\frac{1}{2}}a_1\Big( \dfrac{N}{2\mu} \Big)^{\frac{1}{8}} \Big\}^2 
}
{ 2(1+\pi)a_2 }
\Biggr]
+ \ldots 
\nonumber\\
&&
\label{exprJn2}
\eea
where the dots stand for terms proportional to negative powers of
$N/(2\mu)$.
It is now possible to see how we should scale $\mu$ with $N$. We will
introduce the scaled variable $\nu$ defined by
$2\mu=\nu^4 N^{-\frac{1}{3}}$, where the fourth power on $\nu$ is just a
matter of convenience.
Substituting (\ref{exprJn2}) in (\ref{exprInG}), using the large $z$ expansion
(\ref{calG1aspt}), and neglecting terms that vanish
as $N\to\infty$ we then get
\beq
\expmVo < \mbox{cst}\times N^{\frac{2}{3}}
\int_0^\infty\! \dd\nu\, \ee^{-N^{\frac{1}{3}}g(\nu)}
\label{exprInf}
\eeq
with
\beq
g(\nu) = c_1\nu^{-1} + \frac
{ \big( \tfrac{1}{2}\nu^\frac{1}{2} 
   - a_1\pi^{\frac{1}{2}}\nu^{-\frac{1}{2}} \big)^2 }
{2(1+\pi)a_2}\,.
\label{deff}
\eeq
The function $g(\nu)$ increases as $\sim\nu^{-1}$ for $\nu\to 0$ and as
$\sim\nu$ for $\nu\to\infty$, and hence has a minimum value for some
$\nu=\nu_{\rm min}$. 
Upon doing the integral (\ref{exprInf}) by steepest descent we get
\bea
\expmVo &<& \mbox{cst}\times N^{\frac{1}{2}}
\exp\big[ -N^{\frac{1}{3}}g(\nu_{\rm min}) \big]
\nonumber\\[2mm]
&=& \mbox{cst}\times n^{\frac{1}{4}}\exp\big[ -c_6n^{\frac{1}{6}} \big],
\eea
where $c_6$ is a positive constant.


\section*{Acknowledgments}

HJH kindly thanks Zoltan R\'acz for a useful exchange of correspondence.


\appendix


\begin{thebibliography}{10}

\bibitem{Sylvester1864}
J.J.~Sylvester,
Problem 1491, {\it The Educational Times\,} (April 1864), London.

\bibitem{Valtr95}
P.~Valtr,
{\it Discrete Comput.~Geom.} {\bf 13}, 637 (1995).

\bibitem{Valtr96}
P.~Valtr, 
{\it Combinatorica\,} {\bf 16}, 567 (1996).

\bibitem{Barany99}
I.~B\'ar\'any,
{\it The Annals of Probability\,} {\bf 27}, 2020 (1999).

\bibitem{Hilhorst05a}
H.J.~Hilhorst,
{\it J.~Stat.~Mech.} L02003 (2005).

\bibitem{Hilhorst05b}
H.J.~Hilhorst,
{\it J.~Stat.~Mech.} P09005 (2005).

\bibitem{Hilhorst08a}
H.J.~Hilhorst, {\it Eur.~Phys.~J.~B: Proceedings of Statphys~23,
Genoa, Italy, July 9-13, 2007.} {\it Available online (2008).}

\bibitem{HilhorstCalka08}
H.J.~Hilhorst and P.~Calka,
{\it to appear in J.~Stat.~Phys.} (2008); arXiv:0802.1869.

\bibitem{CalkaSchreiber05}
P.~Calka and T.~Schreiber,
{\it Ann.~Probab.} {\bf 33}, 1625 (2005).

\bibitem{GMOR07}
G.~Gy\"orgyi, N.R.~Moloney, K.~Ozog\'any, and Z.~R\'acz,
{\it Phys.~Rev.~E\,} {\bf 75}, 021123 (2007).

\bibitem{footnote0}
A polygon not enclosing the origin is confined to a half-disk.
The assertion then follows from a result due to J. G. Wendel ({\it Math.~Scand.} {\bf 11}, 109 (1962)). It can also be seen as a consequence of
applying (\ref{asptK}) to a half-disk.

\bibitem{Calka03b}
P.~Calka, 
{\it Adv.~in Appl.~Probab.} {\bf 35}, 863 (2003). 

\bibitem{Schneider93}
R.~Schneider,
{\it Convex bodies: the Brunn-Minkowski theory\,}
(Cambridge University Press, Cambridge, 1993) 

\bibitem{Hilhorst07}
H.J.~Hilhorst,
{\it J.~Phys.~A\,} {\bf 40}, 2615 (2007).

\bibitem{footnote3}
Sums such as $\xi_m=\beta_m+\gamma_m$ and $\eta_m=\gamma_m+\beta_{m+1}$
require caution, since 
the leading $\,\sim\! n^{-1/2}$ behavior of the two terms cancels
and the result is of order $n^{-1}$. 

\bibitem{FORWZ94}
G.~Foltin, K.~Oerding, Z.~R\'acz, R.~Workman, and R.K.P~Zia,
{\it Phys.~Rev.~E\,} {\bf 50}, R639 (1994).

\bibitem{Burkhardt93}
T.W.~Burkhardt, 
{\it J.~Phys.~A} {\bf 26}, L1157 (1993).

\bibitem{MajumdarComtet04}
S.~Majumdar and A.~Comtet, 
{\it Phys.~Rev.~Lett.} {\bf 92}, 225501 (2004). 

\bibitem{MajumdarComtet05} 
S.~Majumdar and A.~Comtet, 
{\it J.~Stat.~Phys.} {\bf 119}, 777 (2005). 

\bibitem{Burkhardtetal07}
T.W.~Burkhardt, G.~Gy\"orgyi, N.R.~Moloney, and Z.~R\'acz,
{\it Phys.~Rev.~E\,} {\bf 76}, 041119 (2007).

\bibitem{KhoshnevisanShi98}
D.~Khoshnevisan and Z.~Shi,
{\it Trans.~Amer.~Math.~Soc.} {\bf 350}, 4253 (1998).

\bibitem{ChenLi03}
X.~Chen and W.V.~Li,
{\it The Annals of Probability\,} {\bf 31}, 1052 (2003).

\bibitem{footnote2}
Our $r(\phi)$ is a factor $\sqrt{3\pi}$ larger than the
function $h$ studied by GMOR \cite{GMOR07}.

\bibitem{Buchta06}
C.~Buchta,
{\it Mathematika\,} {\bf 53}, 247 (2006).

\end{thebibliography}
\end{document}